\documentclass[12pt]{article}
\usepackage[english]{babel}
\usepackage[pdftex]{color}
\usepackage{amsmath}
\usepackage{graphicx}
\usepackage{hyperref}
\usepackage{amsmath}
\usepackage{amsfonts}
\usepackage{amssymb}

\begin{document}

\title{ \textbf{{}Constrained BRST-BFV and BRST-BV Lagrangians for half-integer HS fields on $R^{1,d-1}$}\thanks{Talk presented at SQS'17,  31  July – 05 August, 2017, at JINR, Dubna, Russia}}
\author{\textsc{Alexander A. Reshetnyak\thanks{%
reshet@ispms.tsc.ru}} \\
Laboratory of Computer-Aided Design of Materials, Institute of \\
Strength Physics and Materials Science of SB  RAS, 634055 Tomsk, Russia}
\date{}
\maketitle
\begin{abstract}
Gauge invariant Lagrangian  descriptions of  irreducible and reducible half-integer higher-spin mixed-symmetric massless and massive representations of the Poincare group with off-shell algebraic constraints are constructed within a metric-like formulation in
a $d$-dimensional flat space-time on the basis of a suggested constrained  BRST approach.
A Lorentz-invariant resolution of the BRST complex within  the constrained  BRST formulations produces a gauge-invariant Fang-Fronsdal Lagrangian entirely in terms of the initial triple gamma-traceless  spin-tensor field $\Psi_{(\mu)_{n}}$ with gamma-traceless gauge parameter. The triplet and quartet formulations are derived.  The  minimal (un)constrained  BRST--BV actions for above formulations  are obtained, from proposed  constrained BRST--BV approach  to be by  appropriate tools to construct interacting constrained Lagrangians.
\end{abstract}

\section{Introduction}

Many modern issues of high-energy physics are related to  higher-spin (HS) field theory,
remaining by the part of the LHC experiment program. The  tensionless limit of superstring theory \cite{tensionlessl} with help of BRST operator includes an infinite set of HS fields  with integer and half-integer generalized spins and
incorporates HS field theory into superstring theory and turns it into a method of studying the classical
and quantum structure of the latter (for the present status of HS field theory, see the reviews \cite{reviews}, \cite{reviewsV},   \cite{reviews3}). Whereas  (un)constrained BRST-BFV (see original papers to quantize constrained dynamical systems \cite{BFV}, \cite{BFV1}) and constrained BRST-BV approaches  to construct respectively gauge-invariant Lagrangian formulations (LFs) and BV field-antifield actions for integer HS fields on constant curvature space-times with(out) off-shell holonomic constraints  are known, see e.g. \cite{BRST-BFV1}.  \cite{BRST-BFV2}, \cite{BRST-BFV3}, \cite{BRST-BFV4} and \cite{BRST-BV1}, \cite{BRST-BV2}, \cite{BRST-BV3}  constrained BRST-BFV and BRST-BV methods to be applied to the same aims  for the  half-integer HS fields  have not been till developed.  The paper suggests     constrained BRST-BFV  and  BRST-BV approaches for  construction of  LFs and
 BRST--BV  actions in the minimal sector of the field-antifield formalism for free (ir)reducible Poincare group representations with half-integer
spins  in a flat $\mathbb{R}^{1,d-1}$-space-time subject to an arbitrary Young
tableaux (YT) with $k$ rows, $Y({s}_1,{s}_2,...,{s}_k)$,  with spin $\mathbf{s}=({n}_1+\frac{1}{2},...,{n}_k+\frac{1}{2})$  for $n_1 \geq n_2\geq...\geq n_k$ \cite{metsaevmixirrep}, in a metric-like
formalism (for the study of the LFs in  the metric-like and the frame-like formalisms beyond the BRST methods, see e.g.  \cite{Franciamix} and \cite{Skvortsov}, \cite{Zinoviev}). The latter constrained BRST--BV action presents a natural ground for the procedure of consistent construction of the interacting LFs for such HS fields.

The paper is based on the research \cite{BuchbinderReshetnyakCBFV}, \cite{BuchbinderReshetnyakCBV} and  organized as follows. In Section~\ref{cBRSTBFV}, we suggest  the
constrained BRST--BFV LFs for half-integer   mixed-symmetric (MS) HS fields. In Section~\ref{metric}
a  new formalism is applied   for the case of totally-symmetric (TS), $\mathbf{s}={n}+\frac{1}{2}$,  HS field in various representations.
The construction of a minimal field-antifield  actions on a base of  natural extension of BRST--BFV approach up to a constrained BRST-BV method  for half-integer   HS fields  is considered in Section~\ref{minimalcBRSTBV}.

The convention $\eta_{\mu\nu} = diag (+,
-,...,-)$ for the metric tensor, with the Lorentz indices $\mu, \nu = 0,1,...,d-1$,  the relations $\{\gamma^{\mu},
\gamma^{\nu}\} = 2\eta^{\mu\nu}$ for the Dirac matrices
$\gamma^{\mu}$,and the notation $\epsilon(A)$, $[gh_{H},gh_{L}, gh_{tot}](A)$ for the respective values of Grassmann parity, BFV, $gh_{H}$, BV, $gh_{L}$ and total, $gh_{tot}=gh_{H}+gh_{L}$, ghost numbers of a quantity $A$ are used.  The supercommutator $[A,\,B\}$ of quantities $A, B$
with definite values of Grassmann parity is given by $[A\,,B\}$ = $AB -(-1)^{\epsilon(A)\epsilon(B)}BA$.


\section{Constrained BRST-BFV Lagrangian formulations}

\label{cBRSTBFV} 

There exists two equivalent ways of derivation constrained BRST-BFV approach for LFs for half-integer HS fields \cite{BuchbinderReshetnyakCBFV}, first, from unconstrained BRST-BFV method, developed  for arbitrary  half-integer HS fields in Minkowski space $\mathbb{R}^{1,d-1}$  \cite{Reshetnyk2} (see therein for the references on unconstrained BRST-BFV approach for half-integer HS fields), second, in self-consistent way. We consider here in details the second possibility.    To do so remind, a massless half-integer spin irreducible
representation of the Poincare group in
$\mathbb{R}^{1,d-1}$  is described by a rank-$\sum_{i=1}^k{n}_i $ spin-tensor field
$\Psi_{(\mu^1)_{{n}_1},...,(\mu^k)_{n_k}}
\hspace{-0.2em}\equiv \hspace{-0.2em}
\Psi_{\mu^1_1...\mu^1_{n_1},...,\mu^k_1...\mu^k_{n_k}A}$
with generalized spin $\mathbf{s}$ (for suppressed Dirac index $A$), subject to a YT, $Y(s_1,...,s_k)$ with $k$ rows
of length  $n_1,..., n_k$.
The field  (being symmetric with respect to permutations of each type of Lorentz indices $\mu^i$) satisfies differential (Dirac) equation
(\ref{Eq-0}) and algebraic equations ($\gamma$-traceless and mixed-antisymmetry ones) (\ref{Eq-1}) :
\begin{eqnarray}
\label{Eq-0} &&\hspace{-1.5em}
\imath \gamma^\mu\partial_\mu\Psi_{(\mu^1)_{{n}_1},...,(\mu^k)_{n_k}}
 =0,
\\
&& \hspace{-1.5em}   \gamma^{\mu^i_{l_i}}\Psi_{(\mu^1)_{{n}_1},...,(\mu^k)_{n_k}}=0,
  \quad \Psi_{
(\mu^1)_{n_1},...,\{(\mu^i)_{n_i}\underbrace{,...,\mu^j_{1}...}\mu^j_{l_j}\}...\mu^j_{n_j},...(\mu^k)_{n_k}}=0, \label{Eq-1}
\end{eqnarray}
(for $ 1 \leq l_i \leq n_i, \  1\leq i \leq k$, $i<j$) where the underlined figure bracket  means that the indices inside do not take part symmetrization.

Equivalently, the relations for general state (Dirac spinor)
$|\Psi\rangle $ from  Fock space $\mathcal{H}$ generated by $k$ pairs of bosonic (symmetric case) oscillators
$a^i_{\mu^i}(x), a^{j+}_{\nu^j}(x)$: $[a^i_{\mu^i},
a_{\nu^j}^{j+}]=-\eta_{\mu^i\nu^j}\delta^{ij}$:
\begin{eqnarray}\label{t0t1t}
&&    {t}_0|\Psi\rangle =
{t}_i|\Psi\rangle = t_{rs}|\Psi\rangle =  0, \ g_0^i \Psi\rangle =\textstyle (n_i+\frac{d}{2}) |\Psi\rangle\\
&& \mathrm{for} \ \left({t}_0\,, {t}_i \,,
   t_{rs}\,, g_0^i \right)=\left(-i\tilde{\gamma}^{\mu}\partial_\mu\,,  \tilde{\gamma}_{\mu}a_i^\mu\,,  a^{+}_{r{}\mu}
a_s^{\mu}\,, - \textstyle\frac{1}{2}\{a^{\mu}_i, a^{+}_{i{}\mu}\}\right), \quad {r < s}.\label{totit12}\\
&&
\label{PhysState} |\Psi\rangle =
\sum_{n_1=0}^{\infty}\sum_{n_2=0}^{n_1}\cdots\sum_{n_k=0}^{n_{k-1}}
\frac{\imath^{\sum_in_i}}{n_1!\times...\times n_k!}\Psi_{(\mu^1)_{n_1},...,(\mu^k)_{n_k}}\,
\prod_{i=1}^k\prod_{l_i=1}^{n_i} a^{+\mu^i_{l_i}}_i|0\rangle,
\end{eqnarray}
with the generalized spin  constraints imposed on $|\Psi\rangle$ in terms of number particle operator $g_0^i$ describe
the irreducible massless of spin $\mathbf{s}=\mathbf{n}+\frac{1}{2}$ Poincare group representation. In (\ref{totit12}) we have used  of $d+1$ Grassmann-odd
gamma-matrix-like objects $\{\tilde{\gamma}, \tilde{\gamma}^\mu\}$ to be equivalent to the standard ones: ${\gamma}^{\mu}$,
 whose explicit realization differs in even, $d=2N$, and odd, $d=2N+1$, $N\in \mathbb{N}$ dimensions (see for details \cite{BuchbinderReshetnyakCBFV}).
:For even-valued dimension we have
\begin{eqnarray}
&& \{\tilde{\gamma}^\mu,\tilde{\gamma}^\nu\} = 2\eta^{\mu\nu}, \qquad
\{\tilde{\gamma}^\mu,\tilde{\gamma}\}=0, \qquad
\tilde{\gamma}^2=-1, \label{tgammas} \ \ \mathrm{so}\ \mathrm{that} \ \gamma^{\mu} = \tilde{\gamma}^{\mu} \tilde{\gamma},.
\end{eqnarray}
whereas for odd-valued one, the second and third relations in  (\ref{tgammas}) are changed on   $[\tilde{\gamma}^\mu,\tilde{\gamma}]=0$, $\tilde{\gamma}^2=1$  with unchanged others.
 The set of primary constraints $\{{t}_0, {t}_i, {t}_{ij}, g_0^i\}$, $\epsilon({t}_0)=\epsilon( {t}_i)=1$, $\epsilon({t}_{ij})=\epsilon(g_0^i)=0$ will be closed with respect to the $[\ ,\ \}$--multiplication if we add to them divergentless, $l_i=(1/2)[t_0,t_i\}$, traceless, $l_{ij}=(1/4)[t_i,t_j\}$, $i\leq j$   and D'alamber  operators, $l_0=-t_0^2$:
\begin{equation} \label{lilijpr}
\left(l_0\,,  l_i\,,  l_{ij}\right)=\left( \partial^\mu\partial_\mu\,,- ia_{i{}\mu}
\partial^\mu \,, {\textstyle\frac{1}{2}}\,a_i^{\mu}
a_{j{}\mu}\right).
\end{equation}
The  reality of the Lagrangian  with consistent off-shell holonomic constraints  requires  a closedness for subset of differential constraints, $o_A$,
  with respect to the appropriate hermitian conjugation
 defined by means of odd scalar product in $\mathcal{H}$:
 \begin{eqnarray}
\label{sproduct} \langle\tilde{\Phi}|\Psi\rangle & =  & \int
d^dx\sum_{n_1=0}^{\infty}\sum_{n_2=0}^{n_1}\cdots\sum_{n_k=0}^{n_{k-1}}
         \sum_{p_1=0}^{\infty}\sum_{p_2=0}^{p_1}\cdots\sum_{p_l=0}^{p_{l-1}}\frac{\imath^{\sum_in_i}(-\imath)^{\sum_jp_j}}{n_1!\times...\times n_k!p_1!\times...\times p_l!}
\nonumber\\
&& \times \langle 0|\prod_{j=1}^l\prod_{m_j=1}^{p_j}a^{\nu^j_{m_j}}_j\Phi^+_{(\nu^1)_{p_1},...,(\nu^l)_{p_l}}\tilde{\gamma}_0
\Psi_{(\mu^1)_{n_1},...,(\mu^k)_{n_k}}\,
\prod_{i=1}^k\prod_{l_i=1}^{n_i} a^{+\mu^i_{l_i}}_i|0\rangle\nonumber\\
 &=& \hspace{-0.5em}  \sum_{n_1=0}^{\infty}\sum_{n_2=0}^{n_1}\cdots\sum_{n_k=0}^{n_{k-1}}\prod_{i=1}^k \frac{(-1)^{n_i}}{{n_i!}}\int d^dx
 \Phi^+_{(\mu^1)_{n_1},...,(\mu^k)_{n_k}}\tilde{\gamma}_0
\Psi^{(\mu^1)_{n_1},...,(\mu^k)_{n_k}},
\end{eqnarray}
that means the set: $o_A=\{t_0\,,l_0\,,  l_i\,,l^+_i\}$  (for $l^+_i=-i a^{+}_{i{}\mu}
\partial^\mu$) composes the first-class constraints subsystem. The  holonomic  constraints $t_i$, $t_{rs}$ itself generate the superalgebra of total set of constraints $\{o_a\} = \{t_i, t_{rs}, l_{lm}\}$  in   $\mathcal{H}$. The algebraically independent subset of $\{o_{\bar{a}}\}\subset \{o_a\}$ is given by:
\begin{equation}\label{obara}
   \{o_{\bar{a}}\}  = \{t_1,\, t_{12},\, t_{23},\, t_{34},\,\ldots , t_{(k-1) k}\}.
\end{equation}
As the result the superalgebra  $\mathcal{A}^f_c(Y(k),
\mathbb{R}^{1,d-1})$ to be named as  \emph{constrained half-integer HS symmetry algebra in Minkowski space
with a YT having $k$ rows} with off-shell set of algebraic constraints $\{t_i, t_{rs}\}$  and $g_0^i$ appears by necessary objects to construct constrained LF for  HS fields of spin $\mathbf{s}$. The nilpotent constrained BRST operator for  the system of  $\{o_A\}$ in the Hilbert space $\mathcal{H}_c$,  $\mathcal{H}_c =\mathcal{H}\otimes H^{o_A}_{gh}$, BRST-extended independent algebraic constraints, constrained spin operator determined as
\begin{eqnarray}
 \label{Qc}
 Q_c(o_A) &=&  q_0t_0+\eta_0l_0+\eta_i^+l^i+l^{i+}\eta_i
 + {\imath}\bigl(\sum_l\eta_l^+\eta^l-q_0^2\bigr){\cal{}P}_0,
\\
\label{extconstr}
  \big(\widehat{T}_i,\,\widehat{T}_{rs},\, \widehat{\sigma}{}^i_c(g)\big)  &=&  \big(t_i +  {\mathcal{C}}^A t_{iA}^B \mathcal{P}_B ,\,t_{rs}+{\mathcal{C}}^A t_{rs A}^B \mathcal{P}_B,\, g_0^i+{\mathcal{C}}^A g_{A}^{iB} \mathcal{P}_B\big)+ o(\mathcal{CP}) ,
\end{eqnarray}
with  $gh_H( Q_c, \widehat{T}_i,\,\widehat{T}_{rs},\, \widehat{\sigma}{}^i_c(g))=(1,0,0,0)$, [for $({\mathcal{C}}^A; \mathcal{P}_B)=(q_0,\eta_0,\eta_i^+,\eta_i; p_0, {\cal{}P}_0, {\cal{}P}_j, {\cal{}P}_j^+) $ with $[q_0,p_0]=\{\eta_0, {\cal{}P}_0 \}=\imath$, $\{\eta_i, {\cal{}P}_j^+ \}=\delta_{ij}$]  should satisfy to the \emph{consistency} conditions:
\begin{align}\label{eqQctot}
  & [Q_c,\, \widehat{T}_i\}    = 0, &   [Q_c,\, \widehat{T}_{rs} \}= 0, &&   [Q_c,\, \widehat{\sigma}{}^i_c(g)\}= 0,
\end{align}
known \cite{BuchbinderReshetnyakCBFV} as the \emph{generating equations} for superalgebra of the constrained BRST, $Q_c$ spin operators $\widehat{\sigma}{}^i_c(g)$ and extended  off-shell constraints $\widehat{T}_i,\,\widehat{T}_{rs}$.
   The exact solutions of (\ref{eqQctot}) for unknown  $\widehat{\sigma}{}^i_c(g)$, $\widehat{T}_i,\,\widehat{T}_{rs}$  exists  in the form:
\begin{equation}\label{solextconstr}
 {\mathcal{C}}^A \Big(t_{iA}^B,\, t_{rsA}^B,\, g_{A}^{iB}\Big) \mathcal{P}_B =  \Big(-\imath \eta_ip_0 -2q_0 \mathcal{P}_i ,\, -\eta^+_r\mathcal{P}_s -\mathcal{P}^+_r  \eta_s,\,  \eta_i^+ \mathcal{P}_i- \eta_i \mathcal{P}^+_i
  \Big).
\end{equation}
Presenting the general state vector $|\chi_c\rangle \in \mathcal{H}_c$ (for $({gh}_H, gh_L)|\Psi(a^+_i)^{n_{b{}0} n_{f 0};   (n)_{f i}(n)_{p j}}\rangle = (0,0)$  within the  representation  $(\eta_i, \mathcal{P}_i,  p_0,
\mathcal{P}_0)
|0\rangle=0$):
\begin{eqnarray}
|\chi_c \rangle &=& \sum_n q_0^{n_{b{}0}}\eta_0
^{n_{f 0}} \prod_{i, j}( \eta_i^+ )^{n_{f i}} (
\mathcal{P}_j^+ )^{n_{p j}} |\Psi(a^+_i)^{n_{b{}0} n_{f 0};   (n)_{f i}(n)_{p j}}\rangle \,. \label{chifconst}
\end{eqnarray}
and decomposing  $\mathcal{H}_c$ as $\mathcal{H}_c= \lim_{M\to \infty} \oplus^M_{l=-M}\mathcal{H}^l_c$ for $gh_H(|\chi^l_c \rangle)=-l$, $|\chi^l_c \rangle \in \mathcal{H}^l_c $ from the  BRST condition, $Q_c|\chi^0_c \rangle=0$, off-shell constraints, $\big(\widehat{T}_i,\,\widehat{T}_{rs}\big)|\chi_c \rangle=0$ and  $\widehat{\sigma}{}^i_c(g)|\chi_c \rangle =R^i |\chi_c \rangle$ for some $R^i\in \mathbb{R}$.
we have the   spectral
problem analogous to one for unconstrained case \cite{Reshetnyk2}: %
\begin{align}
\label{Qchic} & Q_c|\chi^l_c\rangle=\delta|\chi^{l-1}_c\rangle, && \widehat{\sigma}{}^i_c|\chi^l_c\rangle=\left( m^i+\textstyle\frac{d-2}{2}\right)|\chi^l_c\rangle,
&& \left(\epsilon, {gh}_H\right)(|\chi^l_c\rangle)=(l+1,-l),
\\
& \big(\widehat{T}_i, \widehat{T}_{rs}\big)|\chi^{l}_c\rangle =0,  &&  l=0,1,...,s_c,   &&\label{constBRST}
\end{align}
with $|\chi^{-1}_c\rangle\equiv 0 $ and $\delta|\chi^{s_c}_c\rangle =0$ for some $s_c$.
 Because of the representation (\ref{chifconst})   the physical state $|\Psi\rangle$ (\ref{PhysState})   is  contained  in $|\chi^0_c\rangle=|\chi_c\rangle= |\Psi\rangle+|\Psi_{Ac}\rangle$ for  $|\Psi_{Ac}\rangle\vert_{\mathcal{C}=0}=0$.

 The system (\ref{Qchic}), (\ref{constBRST}) is compatible, due to the closedness of the superalgebra $\{Q_c,\, \widehat{\sigma}{}^i_c,\, \widehat{T}_i$,  $\widehat{T}_{rs}\}$. Therefore, its resolution   for the joint set of proper eigen-vectors permits one, first, to determine from the middle set and the spin and ghost numbers  distributions for $|\chi^{l}_c\rangle$:
 \begin{eqnarray}\label{nidecomposfconst}
n_i &= & p_i+
n_{f{}i}+n_{p{}i} \,,\  i=1,\ldots,k, \quad
  |\chi^l_c\rangle_{(n)_k}  :\ \  n_{b{}0}+ n_{f{}0}+
     \sum_{i}\bigl(n_{f{}i}- n_{p{}i}\bigr)= -l .
\end{eqnarray}
that $m_i=n_i$ (with $n_i$ from   $\mathbf{s}=\mathbf{n}+\frac{1}{2}$) and proper eigen-vectors $|\chi^{l}_c\rangle_{(n)_k}$.
  The solution of the rest equations is written as the  second-order equations of motion  and sequence of the reducible gauge transformations (\ref{Q12c}) with off-shell constraints (\ref{Q12cc}):
\begin{eqnarray}
&\hspace{-1em}&\hspace{-1em} Q_c|\chi^0_c\rangle_{(n)_k}=0, \ \delta|\chi^0_c \rangle_{(n)_k}
=Q_{c}|\chi^1_c\rangle_{(n)_k} \,, ...\,,
\delta|\chi^{s_c-1}_c \rangle_{(n)_k}
=Q_c|\chi^{s_c}_c\rangle_{(n)_k},\, \delta|\chi^{s_c}_c\rangle_{(n)_k} =0,\label{Q12c}\\
&\hspace{-1em}&\hspace{-1em} \Big(\widehat{T}_i, \widehat{T}_{rs}\Big)|\chi^{l}_c\rangle_{(n)_k} =0, \ \  l=0,1,...,s_c, \ \mathrm{for} \ s_c=
k. \label{Q12cc}
\end{eqnarray}
The corresponding BRST-like  constrained  gauge-invariant action (as for integer HS field)
\begin{equation}\label{2ordercon}
  {\cal{}S}^{(2)}_{c|(n)_k}= \int d \eta_0 {}_{(n)_k}\langle\tilde{\chi}{}^0_c|Q_c |\chi^0_c\rangle_{(n)_k}, \
\end{equation}
  contains second order operator $l_0$, but less terms in comparison with its unconstrained analog \cite{Reshetnyk2}.
  Repeating the procedure of the removing the dependence on $l_0, \eta_0, q_0$ from the BRST operator $Q_{c}$ (\ref{Qc})
and from the whole set of the vectors $|\chi^l_c\rangle_{(n)_k}$ as it was done for unconstrained case  by means of partial gauge-fixing
we come to the:

\noindent
 \textbf{Statement}: The first-order constrained gauge-invariant Lagrangian formulation for half-integer HS field, $\Psi_{(\mu^1)_{n_1},...,(\mu^k)_{n_k}}(x)$ with generalized spin $(s)_k=(n)_k+(\frac{1}{2},...,\frac{1}{2})$, is determined by the action,
  \begin{eqnarray}
\hspace{-0.5em}{\cal{}S}_{c|(n)_k}\hspace{-0.5em} &=&\hspace{-0.5em} \left({}_{(n)_k}\langle\tilde{\chi}^{0}_{0|c}|\  {}_{(n)_k}\langle\tilde{\chi}^{1}_{0|c}|\right)  \left(\hspace{-0.5em}\begin{array}{lr} {t}_0 & \Delta{}Q_c \\
\Delta{}Q_c  &  {t}_0\eta_i^+\eta_i \end{array}
\hspace{-0.5em}\right) \left(\hspace{-0.5em}\begin{array}{c}   | \chi^{0}_{0|c}\rangle_{(n)_k} \\
|\chi^{1}_{0|c}\rangle_{(n)_k}  \end{array}\hspace{-0.5em}
\right) \ \mathrm{for} \  \Delta{}Q_c= \eta_i^+l_i+ l_i^+\eta_i,\label{Lcon}
\end{eqnarray}
  invariant with respect to the sequence of the reducible gauge transformations (for $s_{c}-1=(k-1)$-being by the  the stage of reducibility):
  \begin{eqnarray}
\delta\left(\begin{array}{c}   | \chi^{l(0)}_{0|c}\rangle_{(n)_k} \\
|\chi^{l(1)}_{0|c}\rangle_{(n)_k}  \end{array}
\right) &=& \left(\begin{array}{lr}
\Delta{}Q_c  &  {t}_0\eta_i^+\eta_i \\
{t}_0 & \Delta{}Q_c
\end{array}
\right) \left(\begin{array}{c}   | \chi^{l+1(0)}_{0|c}\rangle_{(n)_k} \\
|\chi^{l+1(1)}_{0|c}\rangle_{(n)_k}  \end{array}
\right) ,  \  \delta\left(\begin{array}{c}   | \chi^{k(0)}_{0|c}\rangle_{(n)_k} \\
|\chi^{k(1)}_{0|c}\rangle_{(n)_k}  \end{array}
\right) =0
\label{GTconst1}
\end{eqnarray}
(for $l =-1,0,...,k-1$ and $| \chi^{-1(m)}_{0|c}\rangle=0$, $m=0,1$)   with off-shell algebraically independent BRST-extended constraints   imposed on the whole set of field and gauge parameters:
\begin{equation}
\widehat{T}_i\Big(|\chi^{l (0)}_c\rangle_{(n)_k}+q_0|\chi^{l (1)}_c\rangle_{(n)_k}\Big)  =0, \  \widehat{T}_{rs}|\chi^{l (m)}_c\rangle_{(n)_k} =0 \  l=0,1,...,k; \ m=0,1 . \label{constralg}
\end{equation}
 The first constraints in $q_0$-independent form read
 \begin{eqnarray}
&\hspace{-1em}&\hspace{-1em}  \begin{array}{lll}
t_i|\chi^{l (0)}_{0|c} \rangle - \eta_i|\chi^{l (1)}_{0|c} \rangle  = 0 ,    &         t_i|\chi^{l (1)}_{0|c}\rangle - 2\mathcal{P}_i|\chi^{l (0)}_{0|c} \rangle  = 0, & \mathcal{P}_i|\chi^{l (1)}_{0|c} \rangle =0
              \end{array} .\label{compoffshelc}
\end{eqnarray}

In the  massive case for $d=2N$, $N\in \mathbb{N}$ we should to add   $k$ pairs of additional even oscillators in differential constraints:  $(L_i,L_i^+)=(l_i+ mb_i,l_i^++ mb_i^+)$ and in  decomposition for $|\chi^{l (1)}_{0|c} \rangle$ (\ref{chifconst}) in accordance with \cite{BuchbinderReshetnyakCBFV}. For odd dimensions, $d=2N+1$ the Lagrangian formulation for massive HS fields may be extrapolated  from one given in even dimension in terms of the ghost-independent or spin-tensor forms, which depends only on the standard Grassmann-even matrices ${\gamma}^\mu$, and does not depend on $\tilde{\gamma}^\mu, \tilde{\gamma}$, due to the presence of the latter only as even degrees inside the Lagrangian, and due to the homogeneity of the gauge transformations w.r.t. $\tilde{\gamma}$ (see \cite{BuchbinderReshetnyakCBFV} for details).
In \cite{BuchbinderReshetnyakCBFV} it is shown the constrained  Lagrangian formulation (\ref{Lcon})--(\ref{constralg}) is equivalent to unconstrained  one for the same HS field and therefore equivalent to the dynamics to be determined  by the initial irreps conditions (\ref{Eq-0}), (\ref{Eq-1}).

\section{Fang-Fronsdal, triplet, quartet   Lagrangians}

\label{metric}
We demonstrate  the above general results  on the example of TS ($k=1$) spin-tensor field, $\Psi_{(\mu)_n}$ of spin $n+\frac{1}{2}$, which is subject to Dirac (\ref{Eq-0}) and  $\gamma$-traceless equation only from (\ref{Eq-1}) and therefore is described
 by 2  Grassman-odd operators ${t}_0 = -\imath\tilde{\gamma}^\mu \partial_\mu$ \,,
$t_1 =  \tilde{\gamma}^\mu a_\mu$ ($a_\mu \equiv  a^1_\mu$) and $g_0^1\equiv g_0 = -\textstyle\frac{1}{2}\{a^{\mu}, a^{+}_{\mu}\}$  acting on  the basic vector
\begin{eqnarray}
\label{PhysStatetot} |\Psi\rangle &=&
\sum_{n=0}^{\infty}
\frac{\imath^{n}}{n!}\Psi_{(\mu)_{n}}\,
a^{+\mu_{1}}\ldots  a^{+\mu_{n}}|0\rangle.
\end{eqnarray}
with respective proper eigen-values $(0,0, (n+\frac{d}{2}))$.

The corresponding  nilpotent constrained BRST operator for the differential first-class $\{{t}_0, l_0,   l_1, l^+_1\}$,  off-shell independent BRST extended constraint $ \widehat{T}_1$ and constrained spin operator $\widehat{\sigma}_c(g)$ in  $\mathcal{H}_c $  have the  form:
\begin{eqnarray}
  && Q_c =   q_0t_0+\eta_0l_0+\eta_1^+l_1+l_1^{+}\eta_1
 + {\imath}\bigl(\eta_1^+\eta_1-q_0^2\bigr){\cal{}P}_0, \label{Qctotsym}
  \\
  &&  \big\{\widehat{T}_1,  \widehat{\sigma}_c(g)\big\}\  = \
  \big\{ t_1-\imath \eta_1p_0 -2q_0 \mathcal{P}_1, \    g_0+ \eta_1^+\mathcal{P}_{1} -\eta_1\mathcal{P}_{1}^+\big\} \label{spinctotsym}
\end{eqnarray}
whose algebra  satisfy to the  relations (\ref{eqQctot}).
The first-order constrained irreducible  gauge-invariant LF for the field, $\Psi_{(\mu)_{n}}$ are given by the relations  (\ref{Lcon}), (\ref{GTconst1})
for $l =-1,0$  with off-shell algebraically independent BRST-extended constraint $\widehat{T}_1$  imposed on the  fields, $| \chi^{m}_{0|c}\rangle_{n}$, $m=0,1$  and gauge parameter $ | \chi^{1(0)}_{0|c}\rangle_{n}$ according to (\ref{constralg}), (\ref{compoffshelc})
for  $| \chi^{1(1)}_{0|c}\rangle_{n} \equiv 0$.
The field vectors and gauge parameter being proper for $\widehat{\sigma}_c(g)$  have the decomposition in ghosts $\eta_1^+, \mathcal{P}_1^+$:
\begin{eqnarray}
   &\hspace{-1.0em}&\hspace{-1.0em} |\chi^{0}_{0|c}\rangle_{n}= |\Psi \rangle_{n} + \eta_1^+\mathcal{P}_1^+ |\chi \rangle_{n-2} = |\Psi \rangle_{n}+ \frac{\imath^{n-2}}{(n-2)! } \eta_1^+\mathcal{P}_1^+ \chi^{(\mu)_{n-2}}\prod_{k=1}^{n-2}a^+_{\mu_k}|0\rangle ,\label{chi0ts}\\
   &\hspace{-1.0em}& \hspace{-1.0em}\big(|\chi^{1}_{0|c}\rangle_{n}\,, |\chi^{1(0)}_{0|c}\rangle_{n}\big) =  \mathcal{P}_1^+ \big(\tilde{\gamma} |\chi_{1} \rangle_{n-1}\,, |\xi\rangle_{n-1}\big)=\frac{\imath^{n-1}}{(n-1)! }\mathcal{P}_1^+ \big(\tilde{\gamma} \chi_1^{(\mu)_{n-1}}\,,\xi^{(\mu)_{n-1}}\big)\prod_{k=1}^{n-1}a^+_{\mu_k}|0\rangle ,\label{chi01ts}\end{eqnarray}
The constraints (\ref{compoffshelc}) are resolved as the $\gamma$-traceless constraint for the gauge parameter  $\xi^{(\mu)_{n-1}}$ and   triple $\gamma$-traceless one for  $\Psi^{(\mu)_n}$:
\begin{eqnarray}
&& \big((t_1)^3|\Psi \rangle_{n}\,,t_1|\xi\rangle_{n-1}\big)= 0, \quad   \tilde{\gamma}|\chi_{1} \rangle_{n-1} = t_1|\Psi \rangle_{n} , \quad   |\chi \rangle_{n-2}= - \textstyle\frac{1}{2}(t_1)^2|\Psi \rangle_{n}\label{gammafield}
\end{eqnarray}
and therefore, $\prod_{i=1}^3\gamma^{\mu_i}\Psi_{(\mu)_{n}} = 0 $, $\gamma^{\mu}\xi_{(\mu)_{n-1}} = 0 $.

The action (\ref{Lcon})) and the gauge transformations in terms of independent  field vector $|\Psi\rangle_n$ in the ghost-free form look as
\begin{eqnarray}
{\cal{}S}_{c|(n)}\hspace{-0.5em} &=&\hspace{-0.5em} {}_{n}\langle\tilde{\Psi}| \hspace{-0.2em} \left(\hspace{-0.2em}{t}_0  - \textstyle\frac{1}{4}(t_1^+)^2{t}_0t_1^2- t_1^+{t}_0t_1 + l^+_1t_1+ t_1^+l_1+\textstyle\frac{1}{2}t_1^+l_1^+t_1^2 +\textstyle\frac{1}{2}(t_1^+)^2l_1t_1
\hspace{-0.2em}\right) \hspace{-0.2em} |\Psi \rangle_{n} ,\label{Lcontotgh}\\
\delta | \Psi\rangle_{n}\hspace{-0.5em} &=&\hspace{-0.5em}  l_1^+ | \xi\rangle_{n-1}.
\label{GTtotsymgh}
\end{eqnarray}
The gauge invariance for the action ${\cal{}S}_{c|(n)}$ is easily checked with use of the Noether  identity:
\begin{eqnarray}
\delta{\cal{}S}_{c|(n)} \frac{\overleftarrow{\delta}}{\delta |\xi\rangle}\hspace{-0.6em}  &=&\hspace{-0.6em} {}_{n}\langle\tilde{\Psi}|\hspace{-0.2em}  \left(\hspace{-0.2em}{t}_0 \hspace{-0.15em} -\textstyle \frac{1}{4}(t_1^+)^2{t}_0t_1^2\hspace{-0.1em}- t_1^+{t}_0t_1\hspace{-0.1em}+ l^+_1t_1\hspace{-0.1em}+ t_1^+l_1\hspace{-0.1em}+\textstyle\frac{1}{2}t_1^+l_1^+t_1^2\hspace{-0.1em} +\textstyle\frac{1}{2}(t_1^+)^2l_1t_1
\hspace{-0.3em}\right) \hspace{-0.2em}l_1^+\hspace{-0.1em} =\hspace{-0.1em}0 ,\label{Noether_identcontotgh}
\end{eqnarray}
modulo the operators $\mathcal{L}(t_1^+,t_0,l_1, l_1^+)t_1$ vanishing  when acting on the  $\gamma$-traceless vectors. The variational derivative of  the functional $\delta {\cal{}S}_{c|(n)} = \langle\tilde{\Psi}| L(t_0,t_1,...)|\xi \rangle + \langle\tilde{\xi}| L^+(t_0,t_1,...)|{\Psi} \rangle $ (with the kernel $L(t_0,t_1,...)$  in (\ref{Noether_identcontotgh})) with respect to the vector $|\xi \rangle$ was introduced above.

 In the spin-tensor  form the  action and the gauge transformations take the familiar form  \cite{Fronsdalhalfint} with accuracy up  to the common coefficient $(n!)^{-1}$:
\begin{eqnarray}
{\cal{}S}_{c|(n)}(\Psi)\hspace{-0.5em} &=&\hspace{-0.5em} (-1)^n \hspace{-0.3em}\int \hspace{-0.2em}d^dx \hspace{-0.1em} \overline{\Psi}{}^{(\nu)_{n}}\hspace{-0.1em}\Big\{\hspace{-0.2em} - \imath \gamma^\mu \partial_\mu  {\Psi}{}_{(\nu)_{n}} \hspace{-0.25em}+ \hspace{-0.1em} \textstyle\frac{1}{4}n(n\hspace{-0.1em}-1)\hspace{-0.1em}\eta_{\nu_{n-1}\nu_{n}} \hspace{-0.1em}(\imath \gamma^\mu \partial_\mu )\eta^{\mu_{n-1}\mu_{n}} {\Psi}{}_{(\nu)_{n-2}\mu_{n-1}\mu_{n}}   \label{LcontotghFronsdal} \\
 &-&  n
\gamma_{\nu_{n}}( \imath \gamma^\mu \partial_\mu )\gamma^{\mu_{n}} {\Psi}{}_{(\nu)_{n-1}\mu_{n}}+ n (\imath \partial_{\nu_n})\gamma^{\mu_{n}} {\Psi}{}_{(\nu)_{n-1}\mu_{n}} +  n ( \imath \partial^{\mu_n})\gamma_{\nu_{n}} {\Psi}{}_{(\nu)_{n-1}\mu_{n}} \nonumber \\
 &-& \hspace{-0.5em}\textstyle \frac{1}{2}n(n-1) \Big(\gamma_{\nu_{n-1}}(\imath \partial_{\nu_n}) \eta^{\mu_{n-1}\mu_{n}} {\Psi}{}_{(\nu)_{n-2}\mu_{n-1}\mu_{n}} + \eta_{\nu_{n-1}\nu_{n}}\gamma^{\mu_{n-1}}(\imath \partial^{\mu_n})  {\Psi}{}_{(\nu)_{n-2}\mu_{n-1}\mu_{n}}\Big)
\hspace{-0.2em}\Big\} ,\nonumber\\
&=&\hspace{-0.5em} (-1)^n \int d^dx  \Big\{ \overline{\Psi}p\hspace{-0.4em}/{\Psi}- \textstyle\frac{1}{4} n(n-1)\overline{\Psi}{}^{\prime\prime}p\hspace{-0.4em}/{\Psi}{}^{\prime\prime}+ n\overline{\Psi}{}^{\prime}p\hspace{-0.4em}/{\Psi}{}^{\prime} -n\overline{\Psi}{}\cdot  p {\Psi}{}^{\prime}- n \overline{\Psi}{}^{\prime} p \cdot {\Psi}{} \label{LcontotghFronsdal1}\\
&+& \textstyle\frac{1}{2} n(n-1)\Big(\overline{\Psi}{}^{\prime} \cdot  p  {\Psi}{}^{\prime\prime} +\overline{\Psi}{}^{\prime\prime} p \cdot  {\Psi}{}^{\prime}\Big)\Big\}, \nonumber\\
\delta {\Psi}^{(\mu)_{n}} &=& - \sum_{i=1}^n\partial^{\mu_i}{\xi}^{\mu_{1}...\mu_{i-1}\mu_{i+1}...\mu_{n}},  \label{gtrFronsdal}
\end{eqnarray}
 where each term in (\ref{LcontotghFronsdal}) and (\ref{LcontotghFronsdal1}) corresponds to the respective summand  in (\ref{Lcontotgh}), whereas for the last expression we have used Fang-Fronsdal  notations \cite{Fronsdalhalfint} with identifications, $p_\mu=-\imath\partial_{\mu}$, $ - \imath \gamma^\mu \partial_\mu=p\hspace{-0.4em}/$ and  $p \cdot {\Psi} =  p_{\mu_1} {\Psi}^{(\mu)_{n}}$.

The triplet formulation to describe  Lagrangian dynamic of the  field ${\Psi}^{(\mu)_{n}}$ with help of the
triplet of spin-tensors $\Psi^{(\mu)_{n}},  \chi_1^{(\mu)_{n-1}},  \chi^{(\mu)_{n-2}}$ and gauge parameter $ \xi^{(\mu)_{n-1}}$ subject to the off-shell 3 constraints on the field vectors, $|\Psi \rangle_{n}$, $|\chi_{1} \rangle_{n-1}$, $|\chi \rangle_{n-2}$ (\ref{gammafield})  and $\gamma$-traceless constraint on $|\xi\rangle_{n-1}$ in the ghost-independent   form
\begin{eqnarray}
&& {\cal{}S}_{c|(n)}(\Psi,\chi_{1} ,\chi )\ =\ {}_{n}\langle\tilde{\Psi}|  {t}_0|\Psi \rangle_{n}  - {}_{n-2}\langle\tilde{\chi}|  {t}_0|\chi \rangle_{n-2}+{}_{n-1}\langle\tilde{\chi}_1| \tilde{\gamma}{t}_0\tilde{\gamma}|\chi_1 \rangle_{n-1} \nonumber \\
&& \phantom{{\cal{}S}_{c|(n)}(\Psi,\chi_{1} ,\chi )}\  -\big({}_{n-1}\langle\tilde{\chi}_1|  \tilde{\gamma}\big\{ l_1|\Psi \rangle_{n}  -l^+_1|\chi \rangle_{n}
 \big\} +h.c.\big) \hspace{-0.2em}  ,\label{Lcontotghtrip}\\
&& \delta \big(| \Psi\rangle_{n}, |\chi \rangle_{n-2},|\chi_{1} \rangle_{n-1}\big)\ =\  \big(l_1^+, l_1, \tilde{\gamma} t_0\big) | \xi\rangle_{n-1}
\label{GTtotsymghtrip}
\end{eqnarray}
coincides with one suggested in \cite{FranciaSagnottitrip}. Without off-shell constraints the triplet formulation describes the  free propagation of couple of massless particles with respective spins $(n+\frac{1}{2})$, $(n-\frac{1}{2})$,...,$\frac{1}{2}$. It was shown in \cite{quartmixbosemas} that this formulation  maybe  described within unconstrained quartet formulation with additional, to the triplet,  compensator field vector $|\varsigma\rangle_{n-2}$, whose gauge transformation  is proportional to  the constraint on $|\xi\rangle_{n-1}$: $\delta|\varsigma\rangle_{n-2}=\tilde{\gamma}t_1|\xi\rangle_{n-1}$ and the whole off-shell constraints (\ref{gammafield}) are augmented by the terms proportional to $|\varsigma\rangle$ to provide theirs total gauge invariance with respect to (\ref{GTtotsymghtrip}) and above gauge transformations for $|\varsigma\rangle$:
\begin{eqnarray}
\hspace{-0.9em}  &\hspace{-0.9em}&\hspace{-0.9em}\big\{ t_1|\Psi \rangle-\tilde{\gamma}|\chi_{1} \rangle+l_1^+\tilde{\gamma}|\varsigma\rangle ,\,   |\chi \rangle+\textstyle\frac{1}{2}t_1\tilde{\gamma}|\chi_{1} \rangle+\textstyle\frac{1}{2}t_0\tilde{\gamma} |\varsigma\rangle ,  t_1 |\chi \rangle + l_1\tilde{\gamma}|\varsigma\rangle\big\} = \big\{0,0,0\big\}\hspace{-0.1em}.\label{gammafieldquart}
\end{eqnarray}
Introducing the respective Lagrangian multipliers: fermionic  ${}_{n-1}\langle\tilde{\lambda}_1 |$, bosonic  ${}_{n-2}\langle\tilde{\lambda}_2 |$,  fermio\-nic  ${}_{n-3}\langle\tilde{\lambda}_3 |$ with trivial gauge transformations,  the equations (\ref{gammafieldquart}) and theirs hermitian conjugated may be derived from the action functional
\begin{eqnarray}\label{Saddquart}
  && {\cal{}S}_{\mathrm{add}|(n)}( \lambda)= {}_{n-1}\langle\tilde{\lambda}_1 |\big(t_1|\Psi \rangle_{n}\hspace{-0.1em}-\tilde{\gamma}|\chi_{1} \rangle_{n-1}+l_1^+\tilde{\gamma}|\varsigma\rangle_{n-2}\big) + {}_{n-2}\langle\tilde{\lambda}_2 |\big(|\chi \rangle_{n-2}\hspace{-0.1em}\\
  && \ \ +\frac{1}{2}t_1\tilde{\gamma}|\chi_{1} \rangle_{n-1}+\frac{1}{2}t_0\tilde{\gamma} |\varsigma\rangle_{n-2}\big)+{}_{n-3}\langle\tilde{\lambda}_3 |\big(t_1 |\chi \rangle_{n-2} + l_1\tilde{\gamma}|\varsigma\rangle_{n-2} \big) + h.c.,\nonumber
\end{eqnarray}
so that, the gauge-invariant functional,
\begin{equation}\label{unconstrStot}
  {\cal{}S}_{(n)}\ = \ {\cal{}S}_{c|(n)}(\Psi,\chi_{1} ,\chi )+ {\cal{}S}_{\mathrm{add}|(n)}( \lambda)
\end{equation}
determines the unconstrained LF for massless spin-tensor of spin $(n+\frac{1}{2})$ in terms of quartet of  spin-tensor fields $\Psi^{(\mu)_{n}},  \chi_1^{(\mu)_{n-1}},  \chi^{(\mu)_{n-2}}, \varsigma^{(\mu)_{n-2}}$ with help of three Lagrangian multipliers $\lambda_i^{(\mu)_{n-i}}$, $i=1,2,3$ with trivial, as it was shown in \cite{quartmixbosemas}, dynamics.

The constrained LF for the massive TS spin-tensor maybe explicitly obtained by means of procedure related to the dimensional reduction \cite{BuchbinderReshetnyakCBFV}.

\section{Constrained Minimal BRST-BV actions}
\label{minimalcBRSTBV}

Here we follow to the research  \cite{BuchbinderReshetnyakCBV}, where the constrained BRST-BV approach to formulate minimal BV action for integer and half-integer HS fields on $\mathbb{R}^{1.d-1}$ is suggested and, in part, to \cite{Reshetnyak_mas} for mixed-antisymmetric integer higher-spin fields. First of all, we weaken the vanishing of $gh_L$ on the component spin-tensors in the decomposition (\ref{chifconst}), when considering instead of field vector $|\chi_c \rangle \in \mathcal{H}_c$ the \emph{generalized field-antifield vector} $|\chi_{g|c} \rangle \in \mathcal{H}_{g|c}  =\mathcal{H}_g\otimes H^{o_A}_{gh}$ with $\mathbb{Z}$-grading for $\mathcal{H}_{g|c}=\lim_{M\to \infty} \oplus^M_{l=-M}\mathcal{H}^l_{g|c}$ for $gh_{\mathrm{tot}}(|\chi^l_{g|c}c \rangle)=-l$, $|\chi^l_{g|c} \rangle \in \mathcal{H}^l_{g|c} $. For simplicity, we consider the case of TS  HS fields.
The total configuration space in the minimal sector  contains (with off-shell constraints) in addition to the triplet  $\Psi_{(\mu)_{n}},  \chi_{1|(\mu)_{n-1}},  \chi_{(\mu)_{n-2}}$  the ghost spin-tensor field $|C^{1(0)}_c\rangle_{n}$ introduced by the rule:
\begin{eqnarray}\label{corrchiC1tot}
 &&  \xi_{(\mu)_{n-1}}(x) =C_{(\mu)_{n-1}}(x) \mu \   \Longrightarrow  \   |\chi^{1(0)}_{0|c}\rangle_{n} =|C^{1(0)}_c\rangle_{n}\mu, \\
 && \mathrm{with} \  (\epsilon, gh_{\mathrm{tot}}, gh_H,gh_L)\big\{C_{(\mu)_{n-1}},\, |C^{1(0)}_c\rangle_{n}\big\}= \big\{(0, 1,0,1), \,  (1,0,-1,1)\big\},\label{ghC1totsym} \end{eqnarray}
which due to the vanishing of the total ghost number and Grassmann parity may be combined with $ |\chi^{0}_c\rangle_{n}$ in  \emph{generalized field vector}:
\begin{eqnarray}\label{genvectortot}
 |\chi^{0}_{\mathrm{gen}|c}\rangle_{n} = |\chi^{0}_c\rangle_{n}+ |C^{1(0)}_c\rangle_{n},  \ \    \big(\epsilon, gh_{\mathrm{tot}}\big)|\chi^{0}_{\mathrm{gen}|c}\rangle=(1,0).
\end{eqnarray}
with untouched $|\chi^{1}_{\mathrm{gen}|c}\rangle_{n}= |\chi^{1}_{0|c}\rangle_{n}$,   $ \big(\epsilon, gh_{\mathrm{tot}}\big)|\chi^{1}_{\mathrm{gen}|c}\rangle=(0,-1)$.

The corresponding   antifield spin-tensors $ {\Psi}^{*(\mu)_{n}},  \chi_1^{*(\mu)_{n-1}},  \chi^{*(\mu)_{n-2}}, C^{*(\mu)_{n-1}}$ with
 \begin{equation}\label{gradtotsym}
 (\epsilon, gh_L){\Psi}^{*}=(\epsilon, gh_L)\chi_1^{*}= (\epsilon, gh_L)\chi^{*}= (0,-1)  \  \mathrm{ and}\  (\epsilon, gh_L)C^* = (1,-2)
 \end{equation}
are combined into \emph{generalized  antifield  vectors}  as follows:
\begin{eqnarray}\label{genavectortot0}
 && |\chi^{*0}_{\mathrm{gen}|c}\rangle_{n} = |\chi^{*0}_c\rangle_{n}+|C^{*1(0)}_c\rangle_{n} = \tilde{\gamma}\Big(|\Psi^{*}(a^+)\rangle_{n} + \mathcal{P}_1^+\eta_1^+ |\chi^{*}(a^+)\rangle_{n-2}\Big)+ \tilde{\gamma}\eta_1^+|C^{*}\rangle_{n-1},    \\
&& |\chi^{*1}_{\mathrm{gen}|c}\rangle_{n} = |\chi^{*1}_c\rangle_{n} = \eta_1^+ |\chi_1^{*}(a^+)\rangle_{n-1},  \ \ \  \big(\epsilon, gh_{\mathrm{tot}}\big)|\chi^{*e}_{\mathrm{gen}|c}\rangle=(1,e-1), e=0,1\label{genavectortot1} \\
&& \ \mathrm{ with }\  \big(gh_L, gh_H\big)|A^{*}\rangle  = (-1,0),\ \mathrm{for} \  A\in\{\Psi, \chi,\chi_1\}, \ \  \big(gh_L, gh_H\big)|C^{*}\rangle = (-2,0) \end{eqnarray}
for the ghost- and $\tilde{\gamma}$- independent antifield vectors $|\Psi^{*}(a^+)\rangle_{n}$, $ |\chi^{*}(a^+)\rangle_{n-2}$, $|\chi_1^{*}(a^+)\rangle_{n-1}$, $|C^{*}\rangle_{n-1}$
having the decompositions in powers of $a^+_\mu$ as for the respective field vectors (\ref{PhysStatetot}), (\ref{chi0ts}), (\ref{chi01ts}) and for the ghost vector $|C(a^+)\rangle_{n-1}$ instead of the  gauge parameter.
The generalized field and antifield vectors (\ref{genvectortot}), (\ref{genavectortot0}), (\ref{genavectortot1}) can be uniquely written in terms of the  generalized field-antifield vector:
\begin{eqnarray}\label{genafvectorts}
 |\chi^{0}_{\mathrm{g}|c}\rangle_{n}& =& \sum\nolimits_{e=0}^1 \Big\{(q_0)^e|\chi^{e}_{\mathrm{gen}|c}\rangle_{n} +(q_0)^{1-e}  |\chi^{*e}_{\mathrm{gen}|c}\rangle_{n}\Big\} , \ \   \big(\epsilon, gh_{\mathrm{tot}}\big)|\chi^{0}_{\mathrm{g}|c}\rangle=(1,0).
\end{eqnarray}
The minimal BV  action for the massless spin-tensor  field ${\Psi}_{(\mu)_{n}}$  in  $\mathbb{R}^{1,d-1}$ takes the form
\begin{eqnarray} \label{Sminconfints}
  S_{c|(n)} & =&  {\cal{}S}_{c|(n)} +    \left\{\left({}_{n} \langle \tilde{\chi}{}^{*0}_c| \left(\eta_1l^+_1+ \eta^+_1l_1\right)+    {}_{n} \langle \tilde{\chi}{}^{*1}_c|t_0 \right) | C^{1(0)}_{c}\rangle_{n}  + h.c.\right\}.
\end{eqnarray}
The functional $S_{c|(n)}$   is invariant with respect to the\emph{ Lagrangian BRST-transformations}
\begin{equation}\label{BRSTSmints}
  \delta_B S_{c|(n)} =0\ \ \mathrm{for} \ \ \delta_B \left(|\chi^{0}_{0|c}\rangle_{n},|\chi^{1}_{0|c}, \rangle_{n}, | C^{1(0)}_{c}\rangle_{n}\right)=  .\mu \left(\eta_1l^+_1+ \eta^+_1l_1,\, t_0, 0\right)| C^{1(0)}_{c}\rangle_{n},
\end{equation}
where the field, antifield vectors are subject to the off-shell BRST extended constraints  according to (\ref{constralg}), (\ref{compoffshelc})
\begin{eqnarray}\label{constralgafts}
&& \widehat{T}_1 \sum\nolimits_{m=0}^1q_0^m |\chi^{e}_c\rangle_{n} = 0 ,\  \ \widehat{T}^*_1\Big(\sum\nolimits_{e=0}^1q_0^{1-e} |\chi^{*e}_c\rangle_{n}\Big)\diagup\{q_0^2\mathcal{P}_1|\chi^{*0}_c\rangle_{n}=0\} =  0,\\
&&
\widehat{T}_1|C^{1 (0)}_c\rangle_{n}=0 ,\ \  \widehat{T}^*_1|C^{*1 (0)}_c\rangle_{n}\diagup\{q_0^2\mathcal{P}_1|\chi^{*1 (0)}_c\rangle_{n}=0\}  = 0.   \label{constralgafts1}
\end{eqnarray}
The resolution of (\ref{constralgafts}), (\ref{constralgafts1}) is reduced due to  (\ref{gammafield}) to the form
\begin{align}
 (t_1)^3\left(|\Psi \rangle, |\Psi^* \rangle\right)= 0, &\qquad   t_1\left(|\Psi \rangle_{n},\,|\Psi^{*} \rangle_{n}\right) = \tilde{\gamma} \left(|\chi_{1} \rangle_{n-1},\,- |\chi^{*}_{1} \rangle_{n-1}\right)  , \label{gammafieldts1}\\
t_1\left(|C_c\rangle_{n-1},\,|C^{*}_c\rangle_{n-1}\right)=0, &\qquad   \left( |\chi \rangle_{n-2},\, |\chi^{*} \rangle_{n-2}\right) =  \textstyle\frac{1}{2}(t_1)^2 \left(-|\Psi \rangle_{n},\, |\Psi^{*} \rangle_{n}\right). \label{gammafieldts2}
\end{align}
In the ghost-independent form the expressions (\ref{Sminconfints}), (\ref{BRSTSmints})  in terms of the triplets of  field $|\Psi \rangle_{n}$, $|\chi_{1} \rangle_{n-1}$, $|\chi \rangle_{n-2}$ and antifield $|\Psi^* \rangle_{n}$, $|\chi^*_{1} \rangle_{n-1}$, $|\chi^* \rangle_{n-2}$ vectors and singlets $|C^{1(0)}_c \rangle_{n-1}$, $|C^{*1(0)}_c \rangle_{n-1}$ (\emph{BV triplet formulation}) read
\begin{eqnarray} \label{bvLcontotghtrip}
 &&  S_{c|(n)}\big(\Psi^{(*)},\chi^{(*)}_{1} ,\chi^{(*)} \big) \ =\  {\cal{}S}_{c|(n)}\big(\Psi,\chi_{1} ,\chi \big)+      \Big\{\Big({}_{n} \langle \tilde{\Psi}{}^{*}| \tilde{\gamma} l^+_1+ {}_{n-2} \langle \tilde{\chi}{}^{*}|\tilde{\gamma}l_1  \\
&&  \phantom{S_{c|(n)}\big(\Psi^{(*)},\chi^{(*)}_{1} ,\chi^{(*)} \big) \ = \ } + {}_{n-1} \langle \tilde{\chi}{}^{*}_1|t_0 \Big) | C\rangle_{n-1}  + h.c.\Big\} , \nonumber \\
&& \delta_B \left(| \Psi\rangle_{n}, |\chi \rangle_{n-2},|\chi_{1} \rangle_{n-1}, |C\rangle_{n-1}\right)\ =\  \mu\left(l_1^+, l_1, \tilde{\gamma} t_0,\,0\right) | C\rangle_{n-1}
\label{bvGTtotsymghtrip},
\end{eqnarray}
and with independent  field $|\Psi\rangle_n$,  ghost $|C^{1(0)}_c \rangle_{n-1}$ and antifield $|\Psi^*\rangle_n$, $|C^{*1(0)}_c \rangle_{n-1}$ vectors
\begin{eqnarray} \label{Sminconfintspsi}
  S_{c|(n)} & =&  {\cal{}S}_{c|(n)}\big(|\Psi\rangle\big) +         \left\{{}_{n} \langle \tilde{\Psi}{}^{*}| \tilde{\gamma} \left(l^+_1+\frac{1}{2} (t_1^+)^2 l_1 - t_1^+t_0 \right) | C\rangle_{n-1}  + h.c.\right\} ,\\
    && \delta_B \left(|\Psi\rangle_{n},\, |C\rangle_{n-1}\right)=  \mu \left(l^+_1,\, 0\right)| C\rangle_{n-1},
\end{eqnarray}
with the classical action given by  (\ref{Lcontotghtrip}) with the constrained (anti)fields (\ref{gammafieldts1}), (\ref{gammafieldts2})    and  (\ref{Lcontotgh}),  $\gamma$-traceless  (anti)field $|C^{(*)}_c\rangle_{n-1}$ and triple $\gamma$-traceless basic (anti)field $| \Psi^{(*)}\rangle_{n}$  in (\ref{Sminconfints}).

On the language of the  component  field ${\Psi}{}_{(\nu)_{n}}$, ghost ${C}{}_{(\nu)_{n-1}}$  spin-tensors  and theirs antifields ${\Psi}{}^{*(\nu)_{n}}$,  ${C}{}^{*(\nu)_{n-1}}$ the constrained minimal BV action and the Lagrangian BRST-transformations take the form in accordance with (\ref{LcontotghFronsdal}), (\ref{gtrFronsdal}) :
\begin{eqnarray}
 && {S}_{c|(n)}(\Psi,C,\Psi^*) = {\cal{}S}_{c|(n)}(\Psi)+  (-1)^n \hspace{-0.3em}
\int \hspace{-0.2em}d^dx \hspace{-0.1em} \Big[\overline{\Psi}{}^{(\nu)_{n}}\hspace{-0.1em}\Big\{ n\Big( \gamma_{\nu_n}   \gamma^\mu \partial_\mu- \partial_{\nu_n}
\Big) C_{(\nu)_{n-1}}  \label{LcontotghFronsdalbv} \\
 && \phantom{{S}_{c|(n)}(\Psi,C,\Psi^*) =} + \textstyle\frac{1}{2}n(n-1)\eta_{\nu_{n-1}\nu_{n}} \partial^{\mu_n}C_{(\nu)_{n-2}\mu_n} \Big\}  +h.c.\Big] \nonumber\\
&&\ =  {\cal{}S}_{c|(n)}(\Psi)+  (-1)^n \int d^dx  \Big\{ n\overline{\Psi}{}^{\prime} p\hspace{-0.4em}/(\imath C) - n\overline{\Psi}\cdot  p (\imath C)  +\textstyle\frac{1}{2} n(n-1)\overline{\Psi}{}^{\prime\prime}p \cdot  (\imath C) +h.c\Big\}, \label{LcontotghFronsdal1bv}\\
&& \delta_B \left({\Psi}^{(\mu)_{n}},\, C^{(\mu)_{n-1}}\right)   \ =\  \mu\big(-  \sum\nolimits_{i=1}^n\partial^{\mu_i} C^{\mu_{1}...\mu_{i-1}\mu_{i+1}...\mu_{n}}, \,0\big). \label{gtrFronsdalbrst}
\end{eqnarray}
The minimal BV action and  Lagrangian BRST-transformations for  unconstrained quartet formulations for massless spin-tensor field are easily obtained from the respective BRST-BFV formulations. E.g. we get for the unconstrained minimal BV action $S_n$:
\begin{eqnarray}\label{unconstrStotbv}
  S_n &=& S_{c|(n)}\big(\Psi^{(*)},\chi^{(*)}_{1} ,\chi^{(*)} \big) +  {\cal{}S}_{\mathrm{add}|(n)}( \lambda)  + \left({}_{n-2}\langle \tilde{\varsigma}{}^*(a)|t_1|C\rangle_{n-1} +h.c. \right) ,\\
  && \delta_B\left(| \Psi\rangle_{n}, |\chi \rangle_{n-2},|\chi_{1} \rangle_{n-1}, |\varsigma \rangle_{n-2},\,| C\rangle_{n-1} \right)\ =\  \mu\left(l_1^+, l_1, \tilde{\gamma} t_0,\,\tilde{\gamma}t_1,\,0\right) | C\rangle_{n-1},
\label{brstunconstrts}
\end{eqnarray}
with account of (\ref{bvLcontotghtrip}), (\ref{unconstrStot}) respectively  for $S_{c|(n)}\big(\Psi^{(*)},\chi^{(*)}_{1} ,\chi^{(*)} \big)$, $ {\cal{}S}_{\mathrm{add}|(n)}( \lambda)$  and antifield vector $ |\varsigma^* \rangle_{n-2}=\tilde{\gamma}|\varsigma^*(a^+) \rangle_{n-2}$. The antifield vectors  in (\ref{unconstrStotbv})  are considered  without off-shell constraints  (\ref{gammafieldts1}), (\ref{gammafieldts2}) as well as the antifield spin-tensors $\lambda^{*(\mu)_{n-i}}_{i}$, $i=1,2,3$, determining the total field-antifield space, do not entered in $S_n$.

The obtained minimal BV actions for TS massless spin-tensor field ${\Psi}{}_{(\mu)_{n}}$ present the basic results of the Section.

The actions  serves to construct  quantum actions under an appropriate choice of a  gauge condition (e.g. for the  TS  field $l_1|\Psi\rangle_n =0$),  as well as to find an interacting theory, including both only the TS half-integer  HS field $\Psi_{(\mu)_{n}}$, with a vertex at least cubic in  $\Psi_{(\mu)_{n}}$ and TS integer HS field $\Phi_{(\mu)_{s}}$., as well as $\Psi_{(\mu)_{n}}$ interacting  with an external electromagnetic field  and some other HS fields.  The consistency of deformation is to be controlled by the master equation for the deformed action with the interaction terms, thus producing a sequence of relations for these terms.

Notice,   the metric-like LF  (\ref{Lcon})--(\ref{compoffshelc})  may be deformed to describe dynamic of  both MS HS field with spin $\mathbf{s}=\mathbf{n}+\frac{1}{2}$  on the AdS(d) space and,  independently, dynamic of MS conformal  HS field  on $\mathbb{R}^{1,d-1}$ which, in turn maybe used to study AdS/CFT correspondence problem.

\vspace{-1ex}

\paragraph{Acknowledgements}

The author is grateful to the organizers of the International Workshop SQS'17 for their hospitality.
I also thank I.L. Buchbinder for the collaboration when solving the problem and to E.D.Skvortsov, D. Francia, A. Campoleoni  for valuable correspondence. The work has been done in the framework of the Program of fundamental research of state academies of Sciences for 2013-2020.

\end{document}